\documentclass[aps,prd,floats,floatfix,showpacs,twoside,preprintnumbers,superscriptaddress,nofootinbib]{revtex4}
\usepackage[dvips]{graphicx,pstricks}
\usepackage{amsmath}
\usepackage{amssymb}
\usepackage{epsfig}
\usepackage{color}
\usepackage[]{hyperref}
\usepackage{dcolumn}

\newcommand{\be}{\begin{equation}}
\newcommand{\ee}{\end{equation}}
\newcommand{\ba}{\begin{eqnarray}}
\newcommand{\ea}{\end{eqnarray}}

\def\roughly#1{\mathrel{\raise.3ex\hbox{$#1$\kern-.75em%
\lower1ex\hbox{$\sim$}}}}

\def\slashchar#1{\setbox0=\hbox{$#1$}  
   \dimen0=\wd0     
   \setbox1=\hbox{/} \dimen1=\wd1  
   \ifdim\dimen0>\dimen1   
      \rlap{\hbox to \dimen0{\hfil/\hfil}} 
      #1     
   \else     
      \rlap{\hbox to \dimen1{\hfil$#1$\hfil}} 
      /      
   \fi}

\makeatletter
\def\overbracket#1{\mathop{\vbox{\ialign{##\crcr\noalign{\kern3\p@}
\downbracketfill\crcr\noalign{\kern3\p@\nointerlineskip}
$\hfil\displaystyle{#1}\hfil$\crcr}}}\limits}
\def\underbracket#1{\mathop{\vtop{\ialign{##\crcr
$\hfil\displaystyle{#1}\hfil$\crcr\noalign{\kern3\p@\nointerlineskip}
\upbracketfill\crcr\noalign{\kern3\p@}}}}\limits}
\def\upbracketfill{$\m@th\makesm@sh{\llap{\vrule\@height3\p@\@width.7\p@}}%
\leaders\vrule\@height.7\p@\hfill
\makesm@sh{\rlap{\vrule\@height3\p@\@width.7\p@}}$}
\def\downbracketfill{$\m@th
\makesm@sh{\llap{\vrule\@height.7\p@\@depth2.3\p@\@width.7\p@}}%
\leaders\vrule\@height.7\p@\hfill
\makesm@sh{\rlap{\vrule\@height.7\p@\@depth2.3\p@\@width.7\p@}}$}
\makeatother

\begin{document}

\date{\today}

\preprint{ZTF-EP-14-01}

\title{Recovering the chiral critical end-point via delocalization
of quark interactions}

\author{S.~Beni\' c\footnote{sanjinb@phy.hr}}
\affiliation{Physics Department, Faculty of Science, 
University of Zagreb, 
Zagreb 10000, Croatia}

\author{D.~Horvati\' c\footnote{davorh@phy.hr}}
\affiliation{Physics Department, Faculty of Science, 
University of Zagreb, 
Zagreb 10000, Croatia}

\author{J.~Klari\' c\footnote{juraklaric@gmail.com}}
\affiliation{Physics Department, Faculty of Science, 
University of Zagreb, 
Zagreb 10000, Croatia}

\begin{abstract}
We show that for the lower branch 
of the quark condensate and values higher than
approximately $-(250 \, \mathrm{MeV})^3$ the 
chiral critical end-point in the Nambu--Jona-Lasinio 
model does not occur in the phase diagram.
By using lattice motivated non-local quark interactions, we 
demonstrate that the critical end-point can be recovered.
We study this behavior for a range of condensate values
and find that the variation in the position of the critical 
end-point is more pronounced
as the condensate is increased.
\end{abstract}


\maketitle

\section{Introduction}

The possibility of a critical end point (CEP) in the 
QCD phase diagram
is a hotly debated issue \cite{Stephanov:2007fk}.
Its speculated existence bears importance for 
heavy ion
collisions, neutron stars and 
perhaps even the early universe.
Since the application of lattice QCD to high chemical potential
leads to the sign problem, the answer will
come from beam energy
scans at RHIC, and the future NICA and FAIR facilities.

Alternatively, models can provide some 
guidance for arguing
the location of the borders in the QCD phase diagram 
and in particular the existence of the CEP, see 
Refs.~\cite{Fukushima:2010bq,Fukushima:2013rx} for reviews.
While in many models one finds the CEP 
\cite{GomezDumm:2005hy,Fukushima:2008wg,
Hell:2008cc,Contrera:2010kz,Contrera:2012wj} (for results
from Dyson-Schwinger 
approach, see \cite{Fischer:2009gk,Qin:2010nq}), 
functional-renormalization
group studies \cite{Herbst:2010rf}, lattice calculations at 
imaginary chemical potential \cite{deForcrand:2006pv}, interplay
with superconductivity \cite{Hatsuda:2006ps} or 
strong vector interaction 
\cite{Sasaki:2006ww} all point that there may be no CEP.

A simple approach to study the chiral phase 
transition and its possible accompanying 
CEP is the 
Nambu--Jona-Lasinio (NJL) model \cite{Nambu:19611,Nambu:19612}.
However, even without its modifications that would
include the vector channel, the diquark channel 
or the Kobayashi-Maskawa-'t Hooft channel 
\cite{Vogl:1991qt,Buballa:2003qv},
the exact position of 
the CEP is rather sensitive 
on the value of the scalar channel coupling.
In fact, as we will demonstrate, if the physical 
coupling is below a 
certain value, the CEP is not present
in the phase diagram.

The intent of this work is to demonstrate that the 
CEP can be restored by delocalizing
the interaction between quarks.
In order to show this we use a 
instantaneous nonlocal variant of the NJL
model \cite{Schmidt:1994di,Blaschke:1994px,Grigorian:2006qe}, see also
\cite{Blaschke:1995gr,Blaschke:2003yn,Grigorian:2003vi,
Aguilera:2006cj,Sasaki:2006ww},
allowing a smooth interpolation 
between highly delocalized and local 
NJL interactions.
The idea of delocalizing quark interactions is well 
motivated by lattice QCD in Landau
\cite{Parappilly:2005ei,Kamleh:2007ud,Schrock:2011hq} 
and in Coulomb gauge \cite{Burgio:2012ph,Burgio:2013mx} 
but also with 
Dyson-Schwinger calculations \cite{Fischer:2006ub,Roberts:2012sv},
\cite{Pak:2011wu} in respective gauges, where 
a strong infrared running
of the quark propagator is observed.

We make a thorough study of the 
dependence of our statement
on the value of quark condensate in vacuum.
Our findings demonstrate that for larger values of 
the condensate, the CEP
is strongly increasing towards higher temperatures 
as the interaction is gradually delocalized.
For smaller values of the condensate
the dependence of the position of the CEP on the 
delocalization of the quark interactions is mild. 

This paper is organized as follows:
in Section \ref{sec:mod} we set up the model and define its 
parameterizations.
The following Section \ref{sec:phase} contains our main results.
In the final Section \ref{sec:conc} we make our conclusions.

\section{Model}
\label{sec:mod}

We work with the $N_f=2$ NJL model where the
delocalized 4-quark interactions are assumed
to have a separable form \cite{Schmidt:1994di,Blaschke:1994px,Grigorian:2006qe}.
The Euclidean action of the model in coordinate space is given as
\be
S_E = \int d^4 x  \left[\bar{q}(-i\slashchar{\partial}+m)q
-\frac{G_S}{2}J_a(x)J_a(x)\right]~,
\label{eq:nnjl}
\ee
with currents
\be
J_a(x)=\int d^4 z \mathcal{F}(z)\bar{q}\left(x+\frac{z}{2}\right)
\Gamma_aq\left(x-\frac{z}{2}\right)~,
\label{eq:crts}
\ee
where $\Gamma_a = (1,i\gamma_5\boldsymbol{\tau})$, 
$\boldsymbol{\tau}$ are
Pauli matrices, $G$ is the interaction strength and $m$ is the current
quark mass.
The interaction parameter is suitably represented 
by a form-factor
$\mathcal{F}(z)$ \cite{Schmidt:1994di}. 
By assuming in addition that the interaction is instantaneous,
i. e. that in momentum space the 
form-factor depends only on the square of the 
three-momenta $\mathcal{F}(\mathbf{p}^2)$, the thermodynamic potential
in the mean-field approximation can be written as
\be
\Omega = \frac{\sigma^2}{2G}-\frac{d_q}{2}\int\frac{d^3 p}{(2\pi)^3}
\left\{E+
T\log\left[1+e^{-\beta(E-\mu)}\right]
+T\log\left[1+e^{-\beta(E+\mu)}\right]
\right\}~,
\label{eq:pot}
\ee
where $\sigma$ is the chiral mass gap, $G$, and $d_q = 2\times 2\times N_c \times N_f$.
The energy of the quark quasi-particle is given as
\be
E(\mathbf{p}) = \sqrt{\mathbf{p}^2 + M^2(\mathbf{p}^2)}~.
\label{eq:disp}
\ee

Delocalization of the quark interactions
has important consequence of yielding
a momentum dependent quark mass $M(\mathbf{p}^2)$
which is a property seen in lattice studies, see 
e.~g. \cite{Burgio:2012ph}.
For the model at hand, the momentum profile
is governed by the form-factor
\be
M(\mathbf{p}^2) = m+\sigma \mathcal{F}(\mathbf{p}^2)~.
\label{eq:}
\ee
The local limit is given as 
$\mathcal{F}(\mathbf{p}^2) = \theta(\Lambda^2 -\mathbf{p}^2)$
where $\Lambda$ is the NJL cutoff.
Therefore, in order to study the influence of the 
delocalized interactions
we use a family of form-factors \cite{Grigorian:2006qe}
\be
\mathcal{F}(\mathbf{p}^2) = \frac{1}{1+\left(\frac{\mathbf{p}}{\Lambda}\right)^{2\alpha}}~,
\label{eq:lor}
\ee
where $\alpha=2$ is the smoothest form-factor that 
can be used and still 
provide convergence of the gap equation, while 
$\alpha \to \infty$ gives
the local NJL limit.

\subsection{Parametrization}

\begin{figure}[t]
\begin{center}
\parbox{15cm}{
\psfig{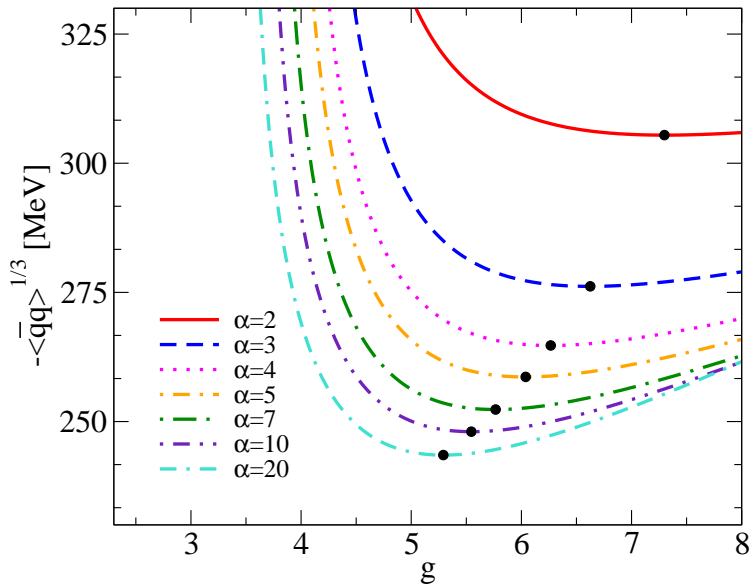}
}
\caption{(Color online) We show the condensate as a function 
$\alpha$ for different reduced couplings $g$.
Dots mark the minimal value separating the lower and the higher
branch, see text.}
\label{fig:cond}
\end{center}
\end{figure}
The parameters of the NJL model $G,\Lambda$ and $m$ are fixed
requiring $m_\pi = 135$ MeV, $f_\pi = 92.4$ MeV and, conventionally
by the vacuum value of the quark 
condensate\footnote{Fixing the constituent 
quark mass $M(0)$, instead of the condensate, is 
another possibility \cite{Grigorian:2006qe,Aguilera:2006cj} which 
we do not consider here.} 
\cite{Grigorian:2006qe}
\be
\langle\bar{q}q\rangle = -2N_c\int\frac{d^3 p}{(2\pi)^3}\frac{M(\mathbf{p})-m}{E(\mathbf{p})}~.
\ee

There are two ambiguities in such a procedure.
The first one is due to the fact that in the instantaneous NJL
there are two values of the
condensate for each coupling, known in the literature as 
the lower and the higher branch \cite{Grigorian:2006qe}, see
Fig.~\ref{fig:cond} where the condensate $\langle\bar{q}q\rangle$
is plotted as a function of the dimensionless coupling
\be
g=G\Lambda^2~,
\ee
by keeping $m_\pi = 135$ MeV and 
$f_\pi = 92.4$ MeV fixed, see Ref.~\cite{Grigorian:2006qe}
for the corresponding equations.
The lower (higher) branch is defined by 
those values of $g$ that lie
one the left (right) from $g$ that gives a minimal $\langle\bar{q}q\rangle$.

We are interested in studying the influence of the parameter $\alpha$
on the CEP.
The large values of $g$ from the higher branch 
are not considered in this work as they yield
large critical temperatures at $\mu=0$ in comparison to 
$T_c(0)\simeq 170$ MeV \cite{Ejiri:2000bw} seen on the lattice.
The family of parametrizations is 
therefore constrained on the lower
branch.
Notice also that in covariant non-local NJL models the higher branch
is absent \cite{GomezDumm:2006vz}.

\begin{table}[htb]
\begin{center}
\begin{tabular}{|c|c|c|c|c|c|c|}
\hline
$\alpha$ & g & $-\langle \bar{q}q\rangle^{1/3}$ [MeV] & 
$\sigma$ [MeV] & $m$ [MeV] & $\Lambda$ [MeV] & $T_c(0)$ [MeV]\\ \hline 
2 & 7.298 & 305.441 & 610.606 & 2.715 & 511.544 & 251.080 \\
3 & 6.625 & 276.165 & 501.450 & 3.660 & 565.332 & 236.659 \\
4 & 6.267 & 264.722 & 467.480 & 4.150 & 579.984 & 232.906 \\
5 & 6.039 & 258.636 & 451.596 & 4.447 & 585.127 & 231.582 \\
7 & 5.766 & 252.329 & 436.747 & 4.786 & 587.916 & 230.920 \\
10 & 5.545 & 248.038 & 426.880 & 5.037 & 588.235 & 230.700 \\
20 & 5.291 & 243.508 & 419.065 & 5.322 & 586.077 & 231.803 \\
\hline
\end{tabular}
\end{center}
\caption{Family of the parameters defined by the minimal condensate for
a particular value of $\alpha$.
The final column contains the respective critical temperatures at $\mu=0$.}
\label{tab:params}
\end{table}
The second ambiguity comes from the value of the chosen quark
condensate, which in general also depends on the renormalization scale.
QCD sum rules provide a value of
$  -(260 \, \mathrm{MeV})^3\lesssim \langle\bar{q}q\rangle 
\lesssim -(190 \, \mathrm{MeV})^3$ 
\cite{Dosch:1997wb}, and the lattice result
$\langle\bar{q}q\rangle(2 \, \mathrm{GeV})^\mathrm{\over{MS}} = 
-(245(4)(9)(7) \, \mathrm{MeV})^3$
from Ref.~\cite{Giusti:1998wy} lies within this range. 
Somewhat higher 
values are supported by recent 
lattice calculation: from Ref.~\cite{Gimenez:2005nt}
we quote 
$\langle\bar{q}q\rangle(2 \, \mathrm{GeV})^\mathrm{\over{MS}} = 
-(265\pm 5\pm 22 \, \mathrm{MeV})^3$,
which is still within the range of sum rules,
while
Ref.~\cite{McNeile:2012xh} finds
$\langle\bar{q}q\rangle = -(283(2) \, \mathrm{MeV})^3$.
With a slight bias towards these higher values we
study a range of 
$-(280 \, \mathrm{MeV})^3 \lesssim \langle\bar{q}q\rangle \lesssim -(240 \, \mathrm{MeV})^3$.

Fig.~\ref{fig:cond} shows that condensate has a higher value
as the interactions are delocalized.
For example, the minimal possible value of the condensate 
with $\alpha=2$ is $\langle\bar{q}q\rangle = -(305.441 \, \mathrm{MeV})^3$ which is outside the said phenomenological range.
Therefore, the most delocalized model that we will use is with
$\alpha=3$ where the minimal condensate 
is $\langle\bar{q}q\rangle = -(276.164 \, \mathrm{MeV})^3$, but still
keep the case $\alpha=2$ as a curiosity\footnote{For example,
by fitting the covariant non-local NJL model to lattice 
Ref.~\cite{Noguera:2008cm} obtained a rather high value 
of $\langle\bar{q}q\rangle = -(326 \, \mathrm{MeV})^3$.
}.

The parametrization of the model is made in the following way: we start
from a particular value of the condensate, which is conventionally
chosen to be exactly the minimal condensate for 
some integer $\alpha_\mathrm{min}$.
For this particular 
condensate we increase $\alpha>\alpha_\mathrm{min}$ along the lower
branch
up to the point where we reach the local limit.
For practical purposes we have observed that $\alpha=50$ is
sufficient.
This procedure is repeated for several values of the condensate,
all conventionally being minimal for some particular integer $\alpha$.
A complete list of minimal values of the 
condensate, along with the full parametrization
of the model, as well as the corresponding results for the critical temperature at zero chemical potential $T_c(0)$, is collected in Table \ref{tab:params}.

\subsection{Critical couplings}
\label{subsec:critcoup}

\begin{figure}[t]
\begin{center}
\parbox{15cm}{
\psfig{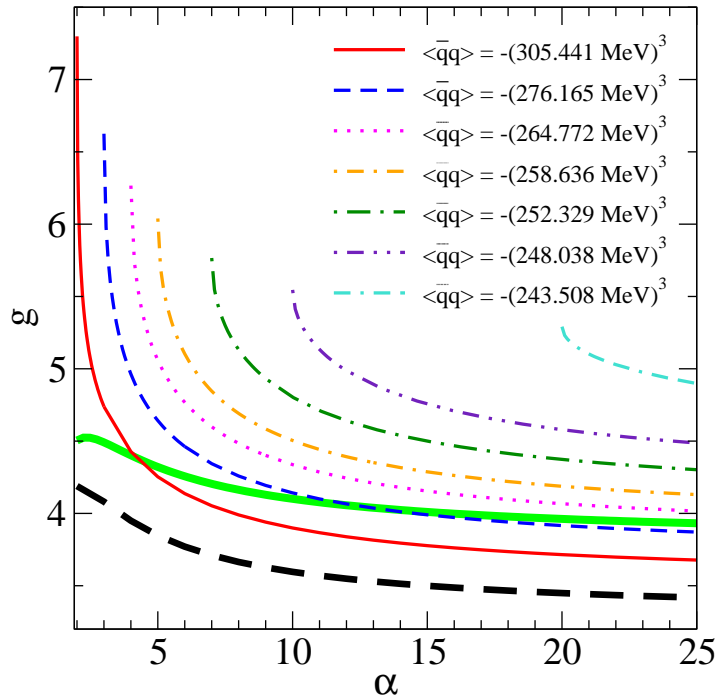}
}
\caption{(Color online) 
We display contours of physical
values of coupling $g$ as functions of $\alpha$, 
indicating the particular values of
the condensate used.
Note that every point on these curves represent a particular
parametrization of the instantaneous NJL model.
The variation of the critical
couplings (dashed, black line) for 
chiral symmetry breaking, and for first order
transition (full, light green line), $g_c$ 
and $\bar{g}_c$, respectively, 
is also shown.}
\label{fig:crit}
\end{center}
\end{figure}

In the limit $m=0$ the
chiral symmetry breaking in the NJL model is established
only for $g>g_c$, where $g_c$ is the critical coupling.
With the delocalized interactions (\ref{eq:lor}) we have
\be
g_c(\alpha) = \frac{8\pi^2}{d_q} \frac{1}{1-\frac{1}{\alpha}}
\frac{\sin\left(\pi/\alpha\right)}{\pi/\alpha}~.
\label{eq:gcrit}
\ee
showing that, for $\alpha>2$, $g_c(\alpha)$ is necessary 
increasing to 
compensate the lack of interaction 
strength from delocalization.
This function is represented by the dashed, black 
curve on Fig.~\ref{fig:crit}.
By increasing $g$ beyond $g_c$ we reach a 
coupling $\bar{g}_c$ where at $T=0$ the
second order transition turns into the first order given by
\be
\bar{g}_c(\alpha) = g_c(\alpha)\left[1-
\left(1-\frac{1}{\alpha}\right)\frac{\sin(\pi/\alpha)}{\pi/\alpha}
(e^{\frac{11}{6}+2\alpha}-1)^{-1/\alpha}\right]^{-1}.
\label{eq:gcritb}
\ee
and shown by the thick, full green line on Fig.~\ref{fig:crit}.
See Appendix \ref{app:critc} for the derivation of (\ref{eq:gcrit})
and (\ref{eq:gcritb}).

While the physical coupling always lies above $g_c$ it does not
necessary lie above $\bar{g}_c$.
On Fig.~\ref{fig:crit} we show contours of physical couplings
along fixed values of $\langle\bar{q}q\rangle$, $f_\pi$ and $m_\pi$,
within a certain range of $\langle\bar{q}q\rangle$.
Even though the physical couplings are not calculated in the chiral
limit it is indicative to observe that for higher values 
of the condensate, $\bar{g}_c$ crosses the physical coupling as
$\alpha$ is increased, i. e. as we proceed to the local limit.
For e.~g. dashed, blue contour, where 
$\langle\bar{q}q\rangle = -(276.164 \, \mathrm{MeV})^3$ this happens
around $\alpha\simeq 15$.

Furthermore, while the physical couplings
at higher values of $\alpha$ increases at roughly 
the same rate as $\bar{g}_c$ by decreasing 
$\alpha$, for smaller values of $\alpha$ it is not so.
In fact, as the form-factor gets more and more delocalized, roughly
in the region $2 \lesssim \alpha \lesssim 10$, the physical
coupling starts to rapidly increase.
This difference between $\bar{g}_c$ and $g$ 
is most severely pronounced for the somewhat 
unrealistic
case of  $\langle\bar{q}q\rangle = -(305.441 \, \mathrm{MeV})^3$,
where Fig.~\ref{fig:crit} shows that $\bar{g}_c$ even drops
a bit at $\alpha=2$.

\section{Phase diagram and the critical end point}
\label{sec:phase}

In this section we study the variation in the position 
of the chiral 
CEP by tuning the non-locality parameter $\alpha$.
We are particularly interested in what happens
for very small values of $\alpha$.
First we find the phase diagram in the chiral limit, for several
values of $\alpha$. 
For physical current mass, and
for several values of $\langle\bar{q}q\rangle$, we employ the
parametrization stated in the previous section and 
calculate the CEP for a range of $\alpha$.

In order to calculate the phase diagram and the CEP we 
first solve the gap equation
\be
\frac{\partial \Omega}{\partial\sigma}=0~,
\ee
and find all possible solutions.
In the case of the 2nd order phase transition (crossover)
there is always one stable and one unstable solution.
The chiral transition
line is found numerically from the divergence (peak) of the thermal
susceptibility $d\sigma / dT$ for the stable solution.
In the case of the 1st order phase transition there are two stable
and one unstable solutions, so the chiral transition is defined by
identifying the global stable solution.
Finally, the CEP is calculated as the point where the unstable solution
observed in the 1st order region merges with the remaining stable solutions.

\subsection{Chiral limit}

\begin{figure}[t]
\begin{center}
\parbox{15cm}{
\psfig{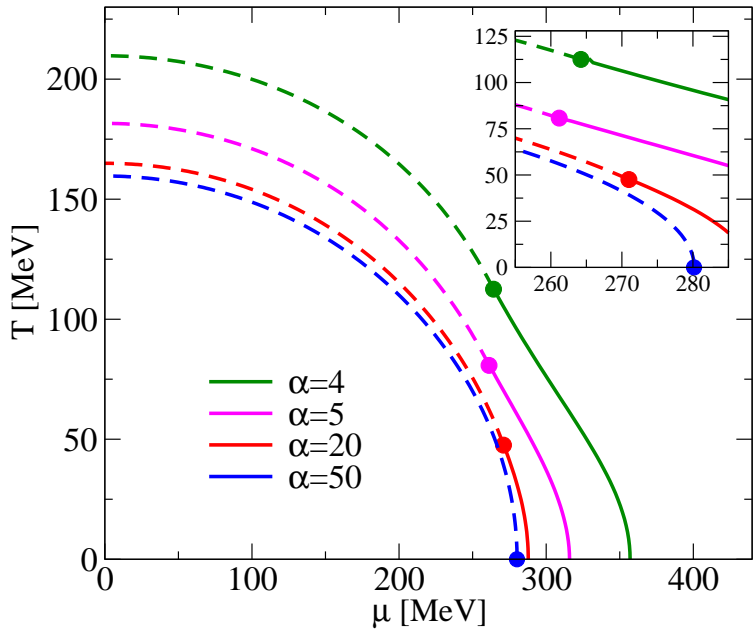}
}
\caption{(Color online) The figure shows several chiral transition curves in the limit $m=0$ 
for $\langle\bar{q}q\rangle = -(265.573 \, \mathrm{MeV})^3$.
We use the parameter sets from table \ref{tab:params2} where we put $m=0$ by hand.
The dashed (full) lines are the second (first) order phase transition.
The case $\alpha=50$, where the CEP
is located at $T=0$, is effectively the local NJL limit.}
\label{fig:pd_chiral}
\end{center}
\end{figure}
In the chiral limit we provide a clean example of 
the impact of the crossing of $\bar{g}_c$
and the physical coupling.
For that purpose we set up a special parametrization where
the physical coupling in the limit $\alpha\to\infty$ is exactly
equal to $\bar{g}_c$ (\ref{eq:gcritb}).
This means that in the local 
NJL limit and the chiral limit, the critical end-point lies exactly
at $T=0$ given by
\be
\mu_c = \Lambda\sqrt{1-\frac{g_c}{g}}~,
\ee
when $g\to \bar{g}_c$ in the NJL limit, see 
Appendix \ref{app:critc}.
The parameterizations of the model are performed for physical quark masses, but the calculation of the phase diagram will be performed in the chiral limit.
The condensate which satisfies the previously stated requirements is 
$\langle\bar{q}q\rangle = -(265.573 \, \mathrm{MeV})^3$.
We then decrease $\alpha$ towards the smoothest possible form-factor allowed
by this particular value of $\langle\bar{q}q\rangle$, which turns out to be
$\alpha=4$.
The relevant results of this particular parametrization procedure are collected in Table \ref{tab:params2}. 

\begin{table}[htb]
\begin{center}
\begin{tabular}{|c|c|c|c|c|}
\hline
$\alpha$ & g & $\sigma$ [MeV] & $m$ [MeV] & $\Lambda$ [MeV]\\ \hline 
4 & 5.799 & 412.066 & 4.111 & 603.352\\
5 & 5.011 & 328.943 & 4.113 & 662.998\\
20 & 4.053 & 261.009 & 4.116 & 746.387\\
50 & 3.903 & 254.578 & 4.117 & 755.169\\
\hline
\end{tabular}
\end{center}
\caption{Family of the parameters for 
$\langle\bar{q}q\rangle = -(265.573 \, \mathrm{MeV})^3$.}
\label{tab:params2}
\end{table}
The chiral transition lines in the limit $m=0$ 
are shown in $\mu-T$ plane
on Fig.~\ref{fig:pd_chiral} for several values of $\alpha$.
Due to our choice of the physical coupling, the 
phase diagram for $\alpha=50$ has a CEP exactly on $T=0$.
Therefore, $\alpha=50$ is an excellent approximation of the local model.
The effect of delocalizing the quark interactions
is that the CEP increases significantly towards non-zero 
temperatures, while the chemical potential of the CEP does not change much.
For the smallest $\alpha$ possible, $\alpha=4$, the CEP 
has a temperature of about $T\simeq 125$ MeV.

Our results are roughly in accordance with the ones shown on 
Fig.~\ref{fig:crit}.
The physical coupling given by the dotted, magenta 
line has almost the same $\langle\bar{q}q\rangle$
as used here, and approaches $\bar{g}_c$, given by the full, green line,
for large values of $\alpha$.
By contrast, decreasing $\alpha$ leads to a large mismatch between the physical
coupling and $\bar{g}_c$, allowing the CEP to significantly increase in the
temperature.

The increase in the critical temperature and 
the chemical potential as $\alpha$ is lowered 
is in part due to the increase in the difference
between $g$ and $\bar{g}_c$, see Fig.~\ref{fig:crit}
but also because the scale
$\Lambda$ is increasing, see Table \ref{tab:params2}.

\subsection{Physical quark masses}

\begin{figure}[t]
\begin{center}
\parbox{15cm}{
\psfig{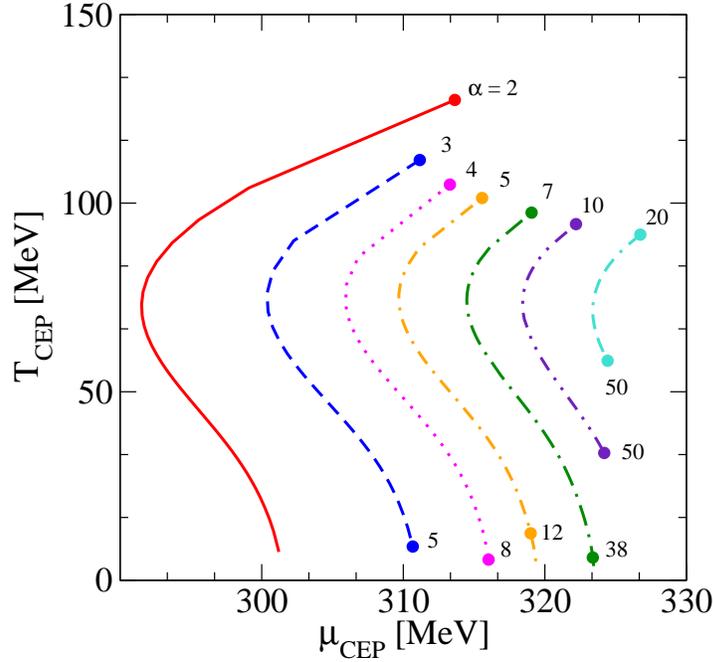}
}
\caption{(Color online) Each curve denotes the position of the CEP as a function of $\alpha$, for a particular value of the
$\langle\bar{q}q\rangle$.
We use the same values of $\langle\bar{q}q\rangle$ and the same line styles 
as defined in Fig.~\ref{fig:crit}.
The values of $\langle\bar{q}q\rangle$ are 
decreased in magnitude as we proceed from
the leftmost to the rightmost curve.
The upper dots indicate the value of the CEP for minimal values
of the condensate, see Table \ref{tab:params}.
The parameter $\alpha$ is varied continuously.
The lower dots indicate the last integer value of $\alpha$
where the CEP occurs in the phase diagram.}
\label{fig:cep}
\end{center}
\end{figure}
At physical quark masses we calculate the CEP for
values of $\langle\bar{q}q\rangle$ defined in \ref{subsec:critcoup}.
Our main result is shown in Fig.~\ref{fig:cep} where 
location of the CEP,
corresponding to these values of $\langle\bar{q}q\rangle$,
of the CEP are shown as a function of $\alpha$, where, starting
from the its minimal value $\alpha$ is varied continuously.
We observe that for several higher values
of $\langle\bar{q}q\rangle$, up to roughly 
$\langle\bar{q}q\rangle \simeq -(250 \, \mathrm{MeV})^3$, the 
CEP vanishes from the phase diagram as $\alpha$ is increased!
Only by delocalizing the quark interactions we are able
to recover CEP in the phase diagram.

Physically, this effect is due to the following.
The 
crossing of 
the physical coupling and $\bar{g}_c$ at large $\alpha$ 
expels the CEP from the phase diagram,
while the large mismatch at low $\alpha$
is responsible for shifting the CEP to high $T$.

For high values of the condensate, such as that shown 
by the dashed, blue line, only the very delocalized
interactions are 
able to hold the CEP in the phase diagram.
Namely, the CEP proceeds rapidly from
$T\simeq 100$ MeV at $\alpha=3$ to $T\simeq 0$ MeV
already for any $\alpha>5$.
It is interesting to observe that the CEP does not
proceed to $T=0$ by reducing both $T$ and $\mu$.
Rather, this happens only for first few values of $\alpha$,
whereas for higher $\alpha$ only $T$ is decreased, while $\mu$
increases.
This effect is also seen in the chiral 
limit, see the inset of Fig.~\ref{fig:pd_chiral}.

In the opposite case, when there is no crossing and the 
physical coupling changes at a similar rate as $\bar{g}_c$,
the CEP is effectively immobilized.
In particular, already for the 
values of $\langle\bar{q}q\rangle = -(243.508 \, \mathrm{MeV})^3$
shown on Fig.~\ref{fig:cep}, the rightmost, cyan line
gives a variation of $\sim 30$ MeV in the temperature.
In such a scenario the CEP is always present.
This is to be expected from the results obtained
in the previous section, and shown in 
Fig.~\ref{fig:crit}, where low values 
of $\langle\bar{q}q\rangle$ do not allow small $\alpha$ 
and thus the physical coupling always lies above $\bar{g}_c$.
Since the actual contours of $g$ shown in Fig.~\ref{fig:cep}
are for physical quark masses, while $\bar{g}_c$ is
obtained in the chiral limit, the values of $\alpha$ at which no CEP occurs in the phase diagram 
is a bit higher than the values of $\alpha$
at which the curves Fig.~\ref{fig:crit} cross $\bar{g}_c$.

Finally, observe that for 
$\langle\bar{q}q\rangle = -(305.441 \, \mathrm{MeV})^3$,
already with $\alpha=3$ no CEP occurs in the  
phase diagram.
The slight offset from the starting points of the 
other families of curves
is attributed to a slight reduction of $\bar{g}_c$ at $\alpha=2$.

\section{Conclusions}
\label{sec:conc}

In this work we have examined how the delocalization of the
quark interactions within the framework of the instantaneous
Nambu--Jona-Lasinio model influences the position of the CEP
in the phase diagram.
Motivated by the lattice calculations \cite{Burgio:2012ph} where the
quark dressing functions, and in particular the mass function, smoothly
changes with momentum we find that the very smooth
form-factors in the instantaneous NJL model 
are possible for the values of the condensate
around $\langle\bar{q}q\rangle \simeq -(280 \, \mathrm{MeV})^3$.
This is somewhat higher than the 
typical values quoted from the 
sum rules \cite{Dosch:1997wb}, but
interestingly, close to a recent 
prediction from the lattice \cite{McNeile:2012xh}.

We show that
delocalization of the quark interactions
drastically influences the position of the CEP.
In particular, there is a gap in the temperature of
$T\sim 100$ MeV between the results in the non-local
with respect to the ones in the 
local model where the CEP tends 
to disappear from the phase diagram.
The minimal value for which this happens, given roughly as 
$\langle\bar{q}q\rangle \simeq -(250 \, \mathrm{MeV})^3$
is still within the range of the values reported from sum rules.
For all higher values the temperature gap
is rather robust to the increase of $\langle\bar{q}q\rangle$.
Lowering the condensate, restricts us to use only rather
local form-factors which in turn immobilize the CEP and
still keep it in the in the phase diagram.

It would be interesting to test further the implications
of the non-local interactions 
on the CEP 
when the full structure of 
the quark propagator,
with the wave function renormalization 
channel taken into account.

\section*{Acknowledgments}
We would like to thank D.~Blaschke and H.~Grigorian for useful discussions.
S.~B. acknowledges the kind hospitality
at the Mini-Symposium on ``Dynamics of Correlations in Dense Hadronic Matter'' in Wroc\l aw.
S.~B. and D.~H. received support by the
University in Zagreb under Contract No.~202348.
This work was supported in part by the COST Action MP1304 ``NewCompStar''.

\begin{appendix}
\section{Critical coupling for first order 
phase transition}
\label{app:critc}

In order to find the critical coupling for which the
CEP in the limit $m=0$ lies exactly at $T=0$ we
make a Landau expansion of the thermodynamic potential
\be
\Omega = \Omega|_{\sigma=0} + \frac{\partial^2\Omega}{\partial\sigma^2}\Big|_{\sigma=0}\sigma^2+
\frac{\partial^4\Omega}{\partial\sigma^4}\Big|_{\sigma=0}\sigma^4+\dots
\ee
where
\be
\begin{split}
\frac{\partial^2\Omega}{\partial\sigma^2}\Big|_{\sigma=0}&=
\frac{\Lambda^2}{g}-\frac{d_q}{2}\int 
\frac{d^3 p}{(2\pi)^3}
\frac{\mathcal{F}^2(\mathbf{p}^2)}{|\mathbf{p}|}
(1-\theta(\mu-|\mathbf{p}|))\\
&=\frac{\Lambda^2}{g}-\frac{\Lambda^2}{g_c}
-\frac{d_q}{16\pi^2}\frac{\mu^2}{\alpha}\left[
\mathcal{F}(\mu^2)+(\alpha-1) _2F_1\left(1,\frac{1}{\alpha},1+\frac{1}{\alpha};
-\left(\frac{\mu}{\Lambda}\right)^{2\alpha}\right)\right]~,
\end{split}
\label{eq:secnd}
\ee
\be
\begin{split}
\frac{\partial^4\Omega}{\partial\sigma^4}\Big|_{\sigma=0}&=
\frac{3d_q}{2}\int\frac{d^3 p}{(2\pi)^3}\left[
\frac{\mathcal{F}^4(\mathbf{p}^2)}{|\mathbf{p}|^3}
(1-\theta(\mu-|\mathbf{p}|)+
\frac{\mathcal{F}^4(\mathbf{p}^2)}{\mathbf{p}^2}
\delta(\mu-|\mathbf{p}|)\right]\\
&=
-\frac{3d_q}{4\pi^2}
\left\{\mathcal{F}^4(\mu^2)+
\frac{1}{6\alpha}\mathcal{F}^3(\mu^2)+
\frac{1}{4\alpha}\mathcal{F}^2(\mu^2)
+\frac{1}{2\alpha}\mathcal{F}(\mu^2)+
\frac{1}{2\alpha}
\log\left[\left(\frac{\mu}{\Lambda}\right)^{2\alpha}
\mathcal{F}(\mu^2)\right]\right\}~.
\end{split}
\label{eq:fourth}
\ee
The function $\mathcal{F}(\mathbf{p}^2)$ is defined 
in Eq.~(\ref{eq:lor}) and $_2F_1(a,b,c;x)$ is the hypergeometric function.
Requiring that  
both (\ref{eq:secnd}) and (\ref{eq:fourth}) vanish 
we find two 
equations for $g$ and $\mu$ defining the CEP.
By assuming $\mu\ll \Lambda$ these yield the critical
chemical potential
\be
\mu_c = \frac{\Lambda}{\left(e^{\frac{11}{6}+2\alpha}
-1\right)^{1/2\alpha}}~,
\ee
and the critical coupling
\be
\bar{g}_c(\alpha) = g_c(\alpha)\left[1-
\left(1-\frac{1}{\alpha}\right)\frac{\sin(\pi/\alpha)}{\pi/\alpha}
(e^{\frac{11}{6}+2\alpha}-1)^{-1/\alpha}\right]^{-1}.
\label{eq:gcritbb}
\ee
In the limit $\alpha\to\infty$ they are given as
\be
\mu_c = \frac{\Lambda}{e}~,
\ee
and
\be
\bar{g}_c = \frac{g_c}{1-e^{-2}}~,
\ee
respectively.

\end{appendix}


\begin{thebibliography}{100}

\bibitem{Stephanov:2007fk} 
  M.~A.~Stephanov,
  PoS LAT {\bf 2006}, 024 (2006)
  [hep-lat/0701002].
  
\bibitem{Fukushima:2010bq} 
  K.~Fukushima and T.~Hatsuda,
  Rept.\ Prog.\ Phys.\  {\bf 74}, 014001 (2011)
  [arXiv:1005.4814 [hep-ph]].
  
\bibitem{Fukushima:2013rx} 
  K.~Fukushima and C.~Sasaki,
  Prog.\ Part.\ Nucl.\ Phys.\  {\bf 72}, 99 (2013)
  [arXiv:1301.6377 [hep-ph]].
  
\bibitem{GomezDumm:2005hy} 
  D.~Gomez Dumm, D.~B.~Blaschke, A.~G.~Grunfeld and N.~N.~Scoccola,
  Phys.\ Rev.\ D {\bf 73}, 114019 (2006)
  [hep-ph/0512218].
  
\bibitem{Fukushima:2008wg} 
  K.~Fukushima,
  Phys.\ Rev.\ D {\bf 77}, 114028 (2008)
  [Erratum-ibid.\ D {\bf 78}, 039902 (2008)]
  [arXiv:0803.3318 [hep-ph]].
  
\bibitem{Hell:2008cc}
  T.~Hell, S.~Roessner, M.~Cristoforetti and W.~Weise,
  Phys.\ Rev.\ D {\bf 79} (2009) 014022
  [arXiv:0810.1099 [hep-ph]].
  
\bibitem{Contrera:2010kz} 
  G.~A.~Contrera, M.~Orsaria and N.~N.~Scoccola,
  Phys.\ Rev.\ D {\bf 82}, 054026 (2010)
  [arXiv:1006.4639 [hep-ph]].
  
\bibitem{Contrera:2012wj} 
  G.~A.~Contrera, A.~G.~Grunfeld and D.~B.~Blaschke,
  arXiv:1207.4890 [hep-ph].
  
\bibitem{Fischer:2009gk} 
  C.~S.~Fischer and J.~A.~Mueller,
  Phys.\ Rev.\ D {\bf 80}, 074029 (2009)
  [arXiv:0908.0007 [hep-ph]].
  
\bibitem{Qin:2010nq}
  S.~-x.~Qin, L.~Chang, H.~Chen, Y.~-x.~Liu and C.~D.~Roberts,
  Phys.\ Rev.\ Lett.\  {\bf 106}, 172301 (2011)
  [arXiv:1011.2876 [nucl-th]].

\bibitem{Herbst:2010rf} 
  T.~K.~Herbst, J.~M.~Pawlowski and B.~-J.~Schaefer,
  Phys.\ Lett.\ B {\bf 696}, 58 (2011)
  [arXiv:1008.0081 [hep-ph]].

\bibitem{deForcrand:2006pv} 
  P.~de Forcrand and O.~Philipsen,
  JHEP {\bf 0701}, 077 (2007)
  [hep-lat/0607017].

\bibitem{Hatsuda:2006ps} 
  T.~Hatsuda, M.~Tachibana, N.~Yamamoto and G.~Baym,
  Phys.\ Rev.\ Lett.\  {\bf 97}, 122001 (2006)
  [hep-ph/0605018].
  
\bibitem{Sasaki:2006ww} 
  C.~Sasaki, B.~Friman and K.~Redlich,
  Phys.\ Rev.\ D {\bf 75}, 074013 (2007)
  [hep-ph/0611147].

\bibitem{Nambu:19611} 
  Y.~Nambu and G.~Jona-Lasinio,
  Phys.\ Rev.\  {\bf 122}, 345 (1961).
  
\bibitem{Nambu:19612} 
  Y.~Nambu and G.~Jona-Lasinio,
  Phys.\ Rev.\  {\bf 124}, 246 (1961).
  
\bibitem{Vogl:1991qt} 
  U.~Vogl and W.~Weise,
  Prog.\ Part.\ Nucl.\ Phys.\  {\bf 27}, 195 (1991).

\bibitem{Buballa:2003qv} 
  M.~Buballa,
  Phys.\ Rept.\  {\bf 407}, 205 (2005)
  [hep-ph/0402234].

\bibitem{Schmidt:1994di} 
  S.~M.~Schmidt, D.~Blaschke and Y.~.L.~Kalinovsky,
  Phys.\ Rev.\ C {\bf 50}, 435 (1994).
  
\bibitem{Blaschke:1994px} 
  D.~Blaschke, Y.~.L.~Kalinovsky, L.~Munchow, V.~N.~Pervushin, G.~Ropke and S.~M.~Schmidt,
  Nucl.\ Phys.\ A {\bf 586}, 711 (1995).
  
\bibitem{Grigorian:2006qe} 
  H.~Grigorian,
  Phys.\ Part.\ Nucl.\ Lett.\  {\bf 4}, 223 (2007)
  [hep-ph/0602238].
  
\bibitem{Blaschke:1995gr} 
  D.~Blaschke, Y.~.L.~Kalinovsky, G.~Roepke, S.~M.~Schmidt and M.~K.~Volkov,
  Phys.\ Rev.\ C {\bf 53}, 2394 (1996)
  [nucl-th/9511003].
  
\bibitem{Blaschke:2003yn} 
  D.~Blaschke, S.~Fredriksson, H.~Grigorian and A.~M.~Oztas,
  Nucl.\ Phys.\ A {\bf 736}, 203 (2004)
  [nucl-th/0301002].
  
\bibitem{Grigorian:2003vi} 
  H.~Grigorian, D.~Blaschke and D.~N.~Aguilera,
  Phys.\ Rev.\ C {\bf 69}, 065802 (2004)
  [astro-ph/0303518].
  
\bibitem{Aguilera:2006cj} 
  D.~N.~Aguilera, D.~Blaschke, H.~Grigorian and N.~N.~Scoccola,
  Phys.\ Rev.\ D {\bf 74}, 114005 (2006)
  [hep-ph/0604196].
  
\bibitem{Parappilly:2005ei} 
  M.~B.~Parappilly, P.~O.~Bowman, U.~M.~Heller, D.~B.~Leinweber, A.~G.~Williams and J.~BZhang,
  Phys.\ Rev.\ D {\bf 73}, 054504 (2006)
  [hep-lat/0511007].
  
\bibitem{Kamleh:2007ud} 
  W.~Kamleh, P.~O.~Bowman, D.~B.~Leinweber, A.~G.~Williams and J.~Zhang,
  Phys.\ Rev.\ D {\bf 76}, 094501 (2007)
  [arXiv:0705.4129 [hep-lat]].
  
\bibitem{Schrock:2011hq} 
  M.~Schrock,
  Phys.\ Lett.\ B {\bf 711}, 217 (2012)
  [arXiv:1112.5107 [hep-lat]].

\bibitem{Burgio:2012ph} 
  G.~Burgio, M.~Schrock, H.~Reinhardt and M.~Quandt,
  Phys.\ Rev.\ D {\bf 86}, 014506 (2012)
  [arXiv:1204.0716 [hep-lat]].

\bibitem{Burgio:2013mx} 
  G.~Burgio, M.~Quandt, H.~Reinhardt and M.~Schrock,
  PoS ConfinementX {\bf }, 075 (2012)
  [arXiv:1301.3619 [hep-lat]].
  
\bibitem{Fischer:2006ub} 
  C.~S.~Fischer,
  J.\ Phys.\ G {\bf 32}, R253 (2006)
  [hep-ph/0605173].

\bibitem{Roberts:2012sv} 
  C.~D.~Roberts,
  arXiv:1203.5341 [nucl-th].
  
\bibitem{Pak:2011wu} 
  M.~Pak and H.~Reinhardt,
  Phys.\ Lett.\ B {\bf 707}, 566 (2012)
  [arXiv:1107.5263 [hep-ph]].
  
\bibitem{Ejiri:2000bw} 
  S.~Ejiri,
  Nucl.\ Phys.\ Proc.\ Suppl.\  {\bf 94}, 19 (2001)
  [hep-lat/0011006].
  
\bibitem{GomezDumm:2006vz}
  D.~Gomez Dumm, A.~G.~Grunfeld and N.~N.~Scoccola,
  Phys.\ Rev.\ D {\bf 74} (2006) 054026
  [hep-ph/0607023].
  
\bibitem{Dosch:1997wb} 
  H.~G.~Dosch and S.~Narison,
  Phys.\ Lett.\ B {\bf 417}, 173 (1998)
  [hep-ph/9709215].
  
\bibitem{Giusti:1998wy} 
  L.~Giusti, F.~Rapuano, M.~Talevi and A.~Vladikas,
  Nucl.\ Phys.\ B {\bf 538}, 249 (1999)
  [hep-lat/9807014].

\bibitem{Gimenez:2005nt} 
  V.~Gimenez, V.~Lubicz, F.~Mescia, V.~Porretti and J.~Reyes,
  Eur.\ Phys.\ J.\ C {\bf 41}, 535 (2005)
  [hep-lat/0503001].
  
\bibitem{McNeile:2012xh} 
  C.~McNeile, A.~Bazavov, C.~T.~H.~Davies, R.~J.~Dowdall, K.~Hornbostel, G.~P.~Lepage and H.~D.~Trottier,
  Phys.\ Rev.\ D {\bf 87}, no. 3, 034503 (2013)
  [arXiv:1211.6577 [hep-lat]].
  
\bibitem{Noguera:2008cm} 
  S.~Noguera and N.~N.~Scoccola,
  Phys.\ Rev.\ D {\bf 78}, 114002 (2008)
  [arXiv:0806.0818 [hep-ph]].
    
\end{thebibliography}
\end{document}